\documentclass[fleqn,10pt]{wlscirep}

\usepackage{amsfonts}
\usepackage{amssymb}
\usepackage{epsfig}
\usepackage{rotating}
\usepackage{lineno}
\usepackage{color}
\usepackage{amsmath}
\usepackage{mathrsfs}
\usepackage{hyperref}
\usepackage{setspace}
\usepackage{grffile}
\usepackage{multirow}
\usepackage{verbatim}   
\usepackage{url}
\usepackage{indentfirst}

\newcommand{\citep}{\cite}
\makeatletter
\def\url@leostyle{%
  \@ifundefined{selectfont}{\def\UrlFont{\sf}}{\def\UrlFont{\small\ttfamily}}}
\makeatother
\usepackage{hyperref}
\hypersetup{pagebackref, bookmarks=true,
citecolor=blue, urlcolor=blue}
\let\epsilon=\varepsilon

\newcommand{\abs}[1]{\left|#1\right|}

\begin{document}

 \title {Increasing Trends of Guillain-Barr\'e Syndrome (GBS) and Dengue in Hong Kong}
\author[1,+]{Xiujuan Tang}
\author[2,+]{Shi Zhao}
\author[2,*]{Alice P.Y. Chiu}
\author[1]{Xin Wang}
\author[3]{Lin Yang}
\author[2,*]{Daihai He}
\affil[1]{Shenzhen Center for Disease Control and Prevention, Shenzhen, China}
\affil[2]{Department of Applied Mathematics, Hong Kong Polytechnic University, Hong Kong}
\affil[3]{School of Nursing, Hong Kong Polytechnic University, Hong Kong}
\affil[*]{daihai.he@polyu.edu.hk and alice.py.chiu@polyu.edu.hk}
\affil[+]{these authors contributed equally to this work}

\parindent 1cm

\begin{abstract}
\section*{Background}
Guillain-Barr\'e Syndrome (GBS) is a common type of severe acute paralytic neuropathy and associated with other virus infections such as dengue fever and Zika. This study investigate the relationship between GBS, dengue, local meteorological factors in Hong Kong and global climatic factors from January 2000 to June 2016.

\section*{Methods}
The correlations between GBS, dengue, Multivariate El Nino Southern Oscillation Index (MEI) and local meteorological data were explored by the Spearman Rank correlations and cross-correlations between these time series. Poisson regression models were fitted to identify nonlinear associations between MEI and dengue. Cross wavelet analysis was applied to infer potential non-stationary oscillating associations among MEI, dengue and GBS.

\section*{Findings}
 An increasing trend was found for both GBS cases and imported dengue cases in Hong Kong. We found a weak but statistically significant negative correlation between GBS and local meteorological factors. MEI explained over 12\% of dengue's variations from Poisson regression models. Wavelet analyses showed that there is possible non-stationary oscillating association between dengue and GBS from 2005 to 2015 in Hong Kong. Our study has led to an improved understanding of the timing and relationship between GBS, dengue and MEI.
\end{abstract}

\flushbottom
\maketitle
\thispagestyle{empty}

\section*{Introduction}
Guillain-Barr\'e Syndrome (GBS) is the most common type of serious acute paralytic neuropathy, with approximately 100,000 new cases worldwide annually \cite{Willison+16}. Approximately two-thirds of these cases are believed to be triggered by prior infections\cite{Wake+13}. GBS has been associated with Zika virus\cite{Pettersson+16}, Chikungunya fever and dengue fever\cite{Willison+16}. GBS cases show peaks in winters than in summers in Western countries \cite{webb2015seasonal}, but not in Latin America and Indian sub-continent \cite{webb2015seasonal}. Previous studies in Hong Kong did not identify any obvious seasonal patterns among adults or children GBS cases \cite{Hui, Ma+10}.

The multivariate El Nino Southern Oscillation index (MEI), is the most comprehensive global index to measure the intensity of El Nino Southern Oscillations (ENSO) \cite{mazzarella2013quantifying}. MEI indicates warm events (``El Nino") from 2014 to present. Previous studies suggested the association between MEI and infectious disease transmission \cite{Hay+00}.

Dengue virus (dengue) is of key public health significance because it can cause rapid and extensive epidemics and thus leads to stresses to the healthcare system \cite{Shepard+16}. Dengue has an estimated 50 million infections per year occurring in approximately 100 endemic countries, including many Southeast Asian countries \cite{WHO+09}.  The global spread of dengue is mainly driven by global trade, increasing travel, urban crowding and ineffective mosquito-control strategies \cite{Simmons+12}, as well as temperature, rainfall and degree of urbanization \cite{Bhatt+13}.

Dengue is a flavivirus, where humans and mosquitoes are the only hosts \cite{Simmons+12}. It is transmitted by Aedes mosquitoes infected with dengue viruses \cite{Simmons+12}. While the principal vector Aedes aegypti is not found in Hong Kong, Aedes albopictus is responsible for the local disease spread. In Hong Kong, over 90\% of the dengue cases are imported cases, i.e. non-locally acquired \cite{Chuang+08}. dengue is mainly found in tropical and sub-tropical countries. They are endemic in many Southeast Asian countries and Southern China \cite{WHO+09}. Previous studies by Tipayamongkholgul et al. and Hurtado-Diaz et al. used autoregressive models to examine the impact of El Nino on dengue incidence \cite{tipayamongkholgul2009effects, hurtado2007short}. A number of wavelet analyses studies had explored the non-stationary oscillating association between dengue and El Nino \cite{vanPanhuis+15, Cazelles2005nonstationary, thai2010dengue, Johansson2009multiyear}. Van Panhuis et al. further reported that there are strong patterns of synchronous dengue transmission across eight Southeast Asian countries. Dengue cycles with a two to five-year periodicity were highly coherent with the Oceanic Nino Index. More synchrony was displayed with increasing temperature \cite{vanPanhuis+15}. Cazelles et al. and Thai et al. also reported on a two to three-year periodicity between dengue and El Nino \cite{Cazelles2005nonstationary}. However, Johansson observed no association between them in Puerto Rico, Mexico and Thailand \cite{Johansson2009multiyear}.

In this work, we aim to study the trends of GBS and dengue in Hong Kong, the association between GBS, dengue and meteorological factors. Wavelet approaches are used to examine the non-stationary oscillating association among these factors.

\section*{Results}
\begin{figure}[!ht]
\centerline{\includegraphics[width=17 cm]{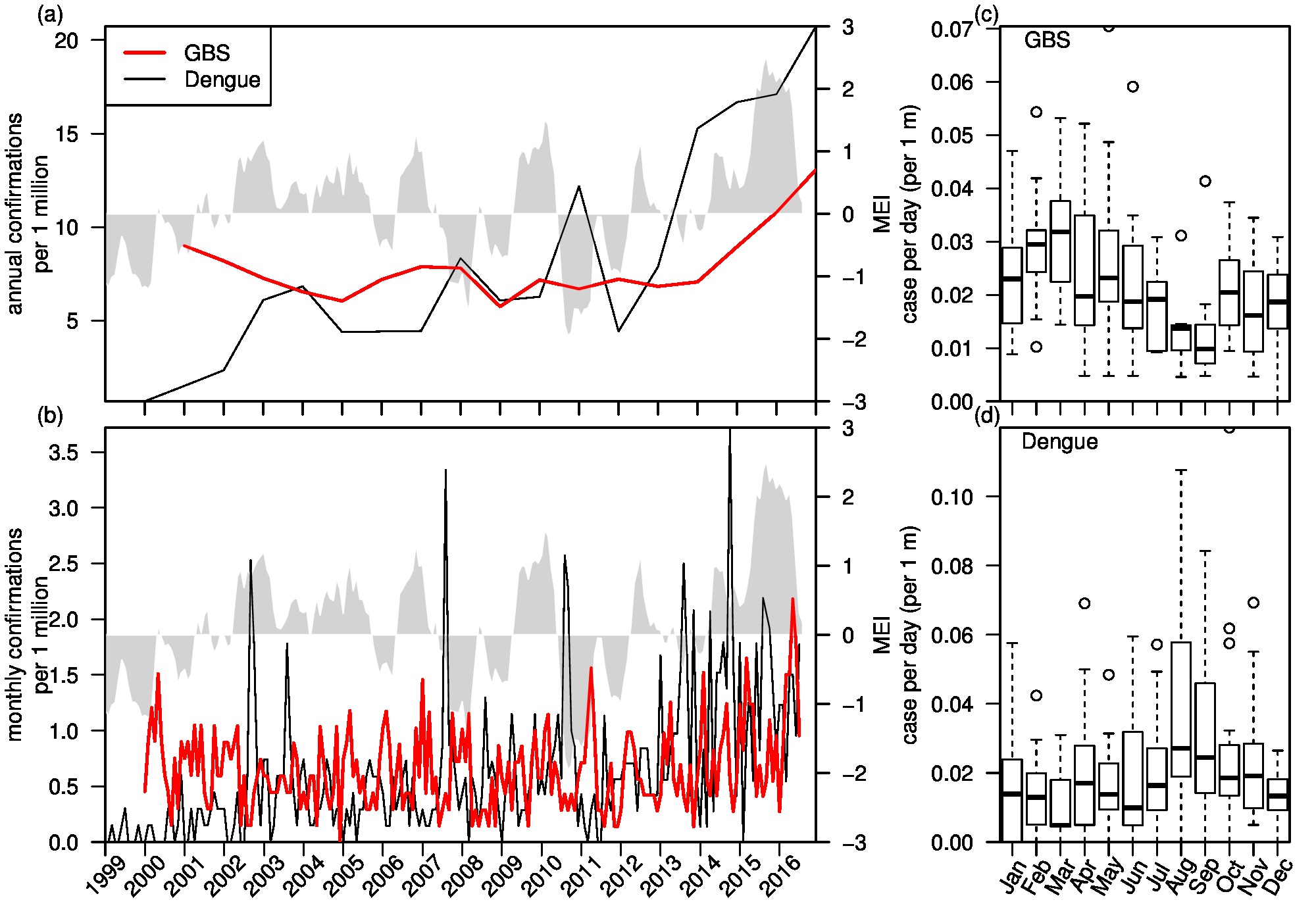}}
\caption{Trends and seasonality of GBS and dengue cases. (a) Annual cases of GBS and dengue cases show a sudden increase in recent years, MEI is represented by the shaded area. (b) Monthly cases of GBS and dengue cases, the shaded area is MEI. (c) Boxplot of GBS cases per day displays a seasonal pattern. (d) Boxplot of dengue cases per day also displays a seasonal pattern.}
\label{fg:mei_gbs_dengue}
\end{figure}

Figure 1 shows the trends and seasonality of GBS and dengue cases in Hong Kong. Annual GBS cases displays mild year-to-year fluctuations, but there is an evident increase after 2014 (Fig 1a). Annual dengue cases also display some variations, but it starts to rise sharply since 2012. Monthly cases of GBS shows mild spikes while monthly dengue cases shows some sharp spikes (Fig 1b). Fig 1c and 1d shows the boxplots of seasonal patterns of GBS and dengue. They display largely opposite seasonal patterns: GBS cases are low in August and September but are high in February and March (Fig 1c); dengue cases are low from February to April but are high in August and September (Fig 1d).

Historical data of MEI presents strong and statistically significant autocorrelations, which can be modelled by an Auto-regressive Model with a time lag of 2 months, i.e. an AR(2) model.

\subsection*{Correlations between GBS and Meteorological Factors}
We first computed correlations between monthly GBS cases and monthly meteorological factors (i.e. median value of daily data in each month) from January 2000 to June 2016. We displayed the results in Table 1. We found weak but statistically significant correlations between mean temperature, minimum temperature, total evaporation with GBS cases. The strongest correlation was about -0.284. We conclude that lower temperature and less evaporation are correlated with more GBS cases in Hong Kong.

\begin{table}[!ht]
\begin{center}
\caption{Correlation ($\rho$) between monthly GBS cases and monthly meteorological factors from January 2000 to June 2016.}
\label{T:cor_gbs_dengue}
\begin{tabular}{c|cccl}
\hline
$\tau$ (months) & $\rho$ & 95\%CI & adjusted $p-$value & signf \\
\hline
1&  -0.029& ( -0.167 , 0.111 )& 1.0000&  \\
2&  0.057& ( -0.082 , 0.195 )& 1.0000&  \\
3&  0.009& ( -0.130 , 0.148 )& 1.0000&  \\
4&  0.173& ( 0.035 , 0.305 )& 0.1713&  \\
5&  0.148& ( 0.009 , 0.281 )& 0.4407&  \\
6&  0.282& ( 0.149 , 0.406 )& 6.371e-04& ***\\
7&  0.292& ( 0.160 , 0.415 )& 3.353e-04& ***\\
8&  0.223& ( 0.087 , 0.352 )& 0.0182& *\\
9&  0.109& ( -0.031 , 0.244 )& 1.0000&  \\
10&  0.120& ( -0.019 , 0.255 )& 1.0000&  \\
11&  0.093& ( -0.047 , 0.229 )& 1.0000&  \\
12&  -0.015& ( -0.154 , 0.125 )& 1.0000&  \\
\hline
\end{tabular}

\end{center}
\end{table}

Mean pressure
Maximum temperature
Mean temperature
Minimum temperature
Mean dew point
Mean relative humidity
Mean amount of cloud
Total bright sunshine
Global solar radiation
Total evaporation
Prevailing wind direction

\subsection*{Cross-correlations of GBS and dengue}
Hypothesizing that dengue could have an effect on GBS with a time lag of several months, we computed the cross-correlation between GBS and dengue cases with different time lags (Table 2). We found that dengue cases do not significantly correlate with GBS, unless a time lag is introduced. The cross-correlation coefficient attains a maximum at 0.2753 (95\% CI: 0.1391, 0.4012) with a time lag of seven months. Our results suggest that dengue could have an effect on GBS with a time delay.

\subsection*{Cross-Correlations of dengue and GBS with MEI}
We computed the cross correlations between dengue and MEI, and between GBS and MEI (Figure 2). The maximum cross-correlation is attained when there is a positive time lag of about three to five months in both GBS and dengue. Panel (a) shows that dengue is positively correlated with MEI, and the correlation is about 0.3 when the time lag is three months, which is biologically reasonable. However, panel (b) suggests that the cross-correlation between GBS and MEI is weaker (maximum correlation is about 0.2) than between dengue and MEI,
These results are consistent with several studies that MEI played a crucial role on mosquito-borne diseases including dengue  \cite{vanPanhuis+15, Cazelles2005nonstationary, thai2010dengue}, but possibly less related to GBS.

\begin{table}[!ht]
\begin{center}
\caption{Cross-correlation ($\rho$) between GBS and dengue from January 2000 to June 2016. Time lag ($\tau$, in months) indicates the number of months GBS lags behind dengue.}
\begin{tabular}{r|rrcl}
\hline
factor& $\rho$& 95\%CI& adjusted $p$-value& signif\\
\hline
mean.pressure&  0.209& ( 0.072 , 0.338 )& 0.0336& *\\
temperature.max&  -0.220& ( -0.348 , -0.083 )& 0.0198& *\\
temperature.mean&  -0.237& ( -0.364 , -0.102 )& 0.0081& **\\
temperature.min&  -0.244& ( -0.371 , -0.109 )& 0.0056& **\\
mean.dew.point&  -0.199& ( -0.329 , -0.061 )& 0.0541& *\\
mean.relative.humidity&  0.124& ( -0.016 , 0.258 )& 0.8987&  \\
mean.amount.of.cloud&  0.139& ( 0.000 , 0.273 )& 0.5495&  \\
total.bright.sunshine&  -0.238& ( -0.365 , -0.102 )& 0.0079& **\\
daily.global.solar.radiation&  -0.188& ( -0.319 , -0.050 )& 0.0853& *\\
total.evaporation&  -0.284& ( -0.407 , -0.151 )& 5.200e-04& ***\\
prevailing.wind.direction&  -0.176& ( -0.308 , -0.038 )& 0.1413&  \\
\hline
\end{tabular}

\label{T:cor_gbs_weather}
\end{center}
\end{table}

\begin{figure}[!ht]
\centerline{\includegraphics[width=15.5cm, height=7.5cm]{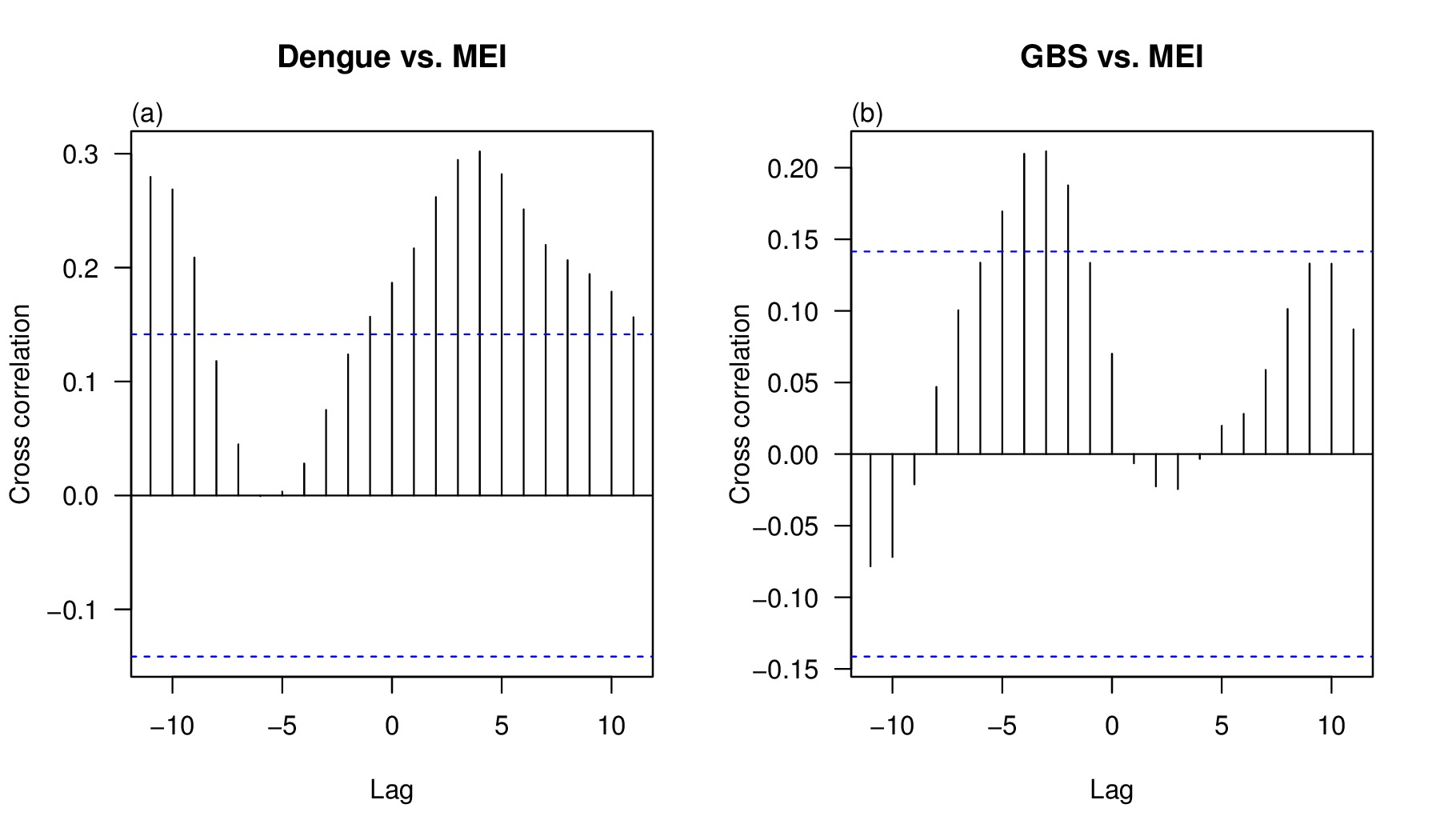}}
\caption{(a) Cross-correlation coefficient between dengue and MEI and (b) Cross-correlation coefficient between MEI and GBS. In both panels, we consider time lag of 0 to 11 months. Dashed lines on both panels indicate the significance level.
} \label{fg:ccf_mei_gbs_dengue}
\end{figure}

We found that dengue is significantly correlated with MEI for the period January 1999 to June 2016 when we did not consider time lag. The cross-correlation coefficient is 0.2048 (95\%CI: 0.0694, 0.3328; p-value=0.003295).

Furthermore, we applied a Poisson regression model to estimate the association between dengue and MEI (eqn. 1), adopting a Moving Average (MA) model due to the relatively large oscillation of dengue data. The results of MA(3) and MA(5) models of dengue are reported in Fig. 3.


\begin{figure}[!ht]
\centerline{\includegraphics[width=17cm, height = 7cm]{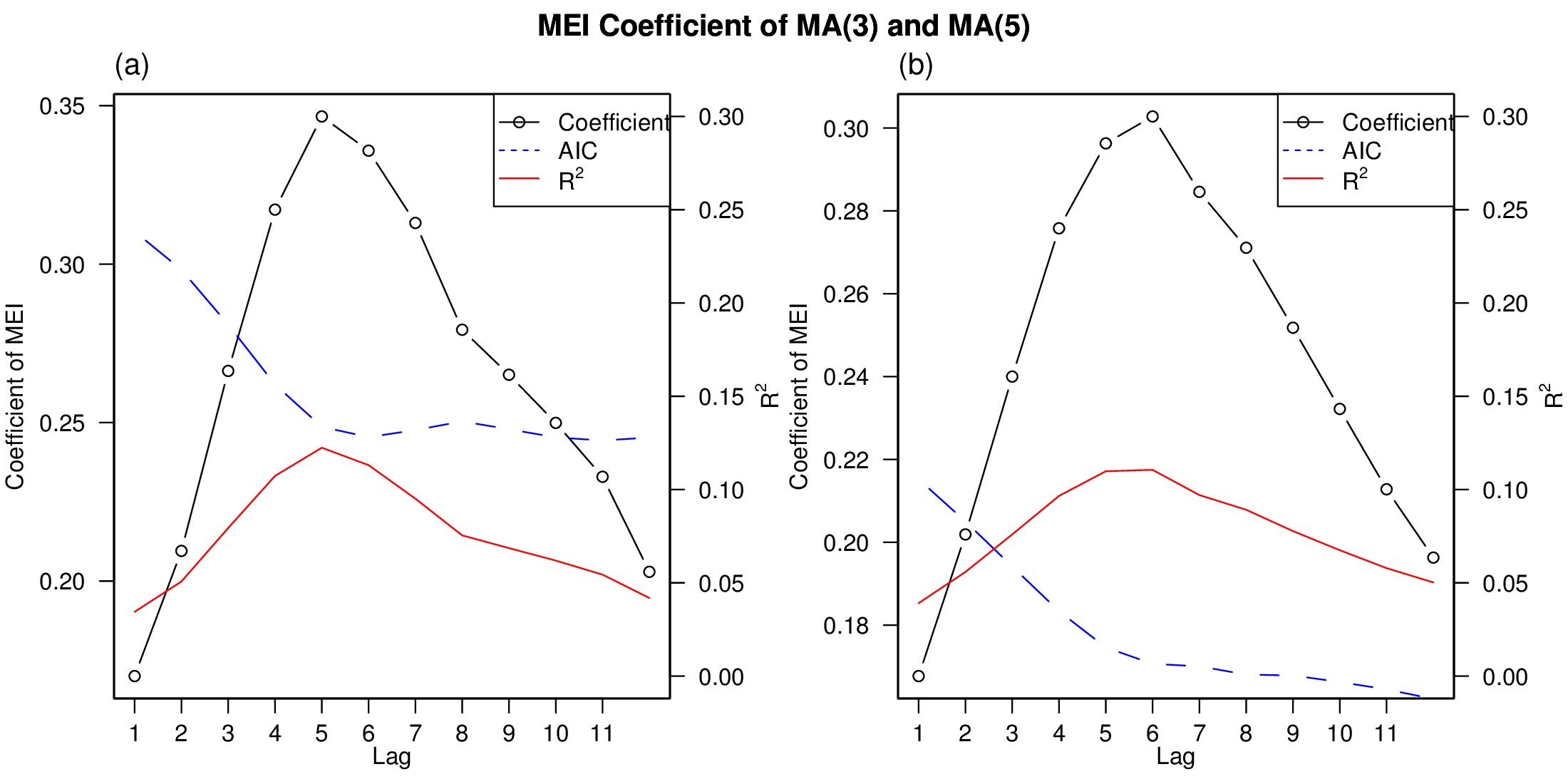}}
\caption{(a) MA(3) model of dengue (b) MA(5) model, displaying lags from 0 to 11 months. Black, dotted: Poisson regression coefficient of MEI (beta); red: $R^2$; blue, dashed: scaled AIC.}
\label{fg:mei_dengue_poisson_regression}
\end{figure}

MEI coefficients (beta) and $R^2$ attain maximum when the lag are in between four to seven months (Figure 3). For both MA(3) and MA(5), the maximum $R^2$ are attained at lags 4 and lag 5, where they are greater than 0.11 (Appendix Table 3 and Table 4). Both results are consistent with previous studies \cite{tipayamongkholgul2009effects,hurtado2007short}.

\subsection*{Wavelet Analyses on MEI, dengue and GBS}
In Figure 4(a), the wavelet transform suggest that MEI are significant at two to three-year periodic band. Figures 4(b) and 4(c) show that dengue and GBS display similar modes, both wavelet power spectrums are significant at around one-year periodic band.

The cross wavelet analyses reveal considerable and significant coherence between MEI and both dengue and GBS since 2010, comparing with
the situation before 2010 (see Fig.~5), which is similar with \cite{Cazelles2005nonstationary,Johansson2009multiyear}.
The periods are at 1-2 years bond for MEI vs. dengue and 0.75-1.75 years bond for MEI vs. GBS from 2000-15.

Figure 6 suggests the oscillation mode (between dengue and GBS) is sometimes in a tone  with a periodic bond of 0.5-1.5 years from 2000-15 in Hong Kong,
the association became significant in the recent year.
Interestingly, the appearance of the significant association between dengue and GBS seemingly coincided with major dengue outbreaks as shown in
Fig. 6a-b.

\begin{figure}[!ht]
\centerline{\includegraphics[width=16.5cm,height = 18cm]{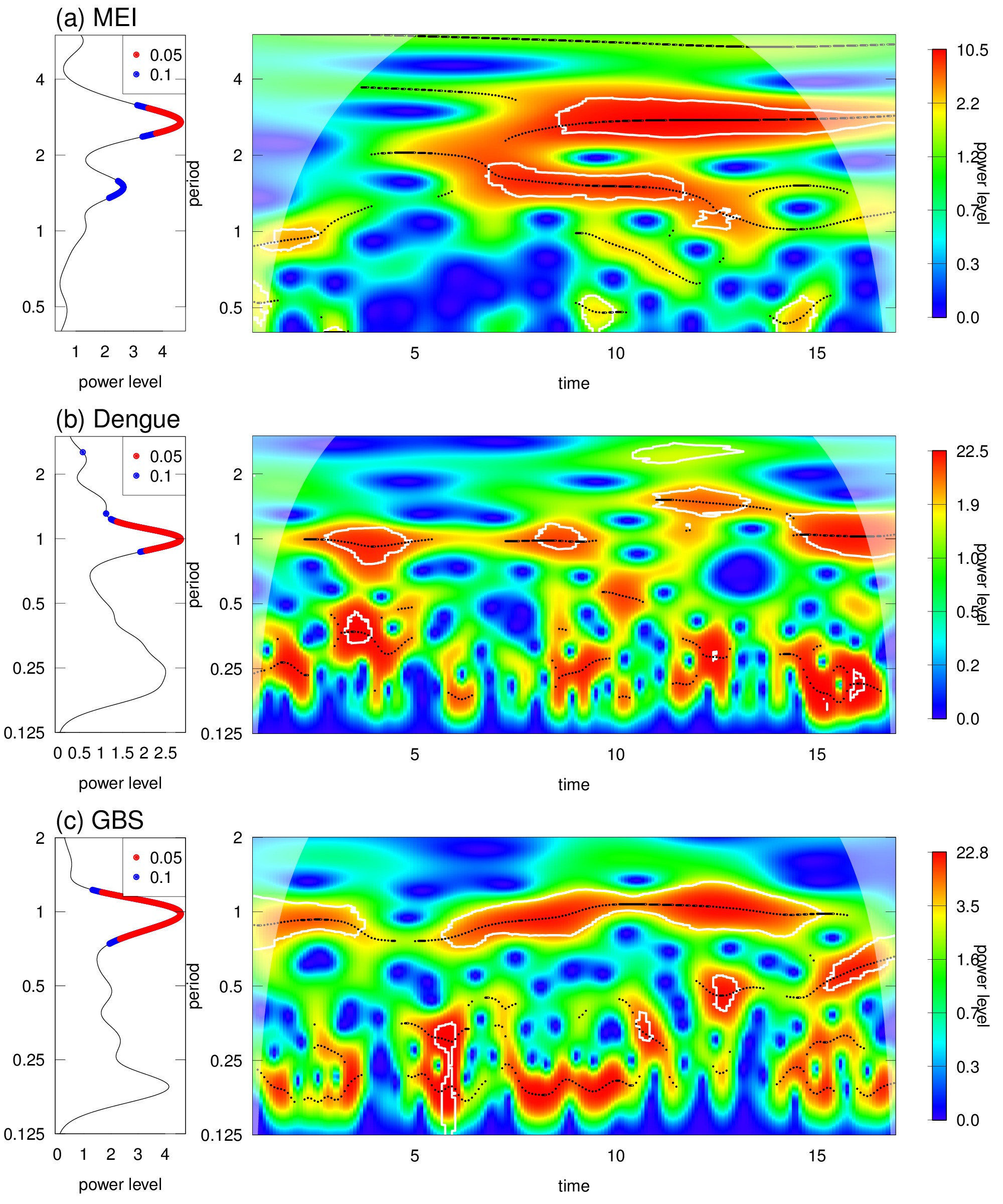}} 
\caption{Wavelet analyses of MEI, dengue and GBS from 2000-2016 in panels (a), (b), and (c). Left panel: mean spectrum plots at 5\% (blue) and 10\% (red) thresholds. (ii) the right panel shows the wavelet power spectrum contour plots. The colour scheme is from blue to red, which represents increasing power level. The white line represents the 95\% C.I. and the white shaded region is due to the edge effect. All data are transformed by taking square roots.}
\label{fg:wavelet_mei_dengue_gbs}
\end{figure}

\begin{figure}[!ht]
\centerline{\includegraphics[width=16.5cm,height = 15cm]{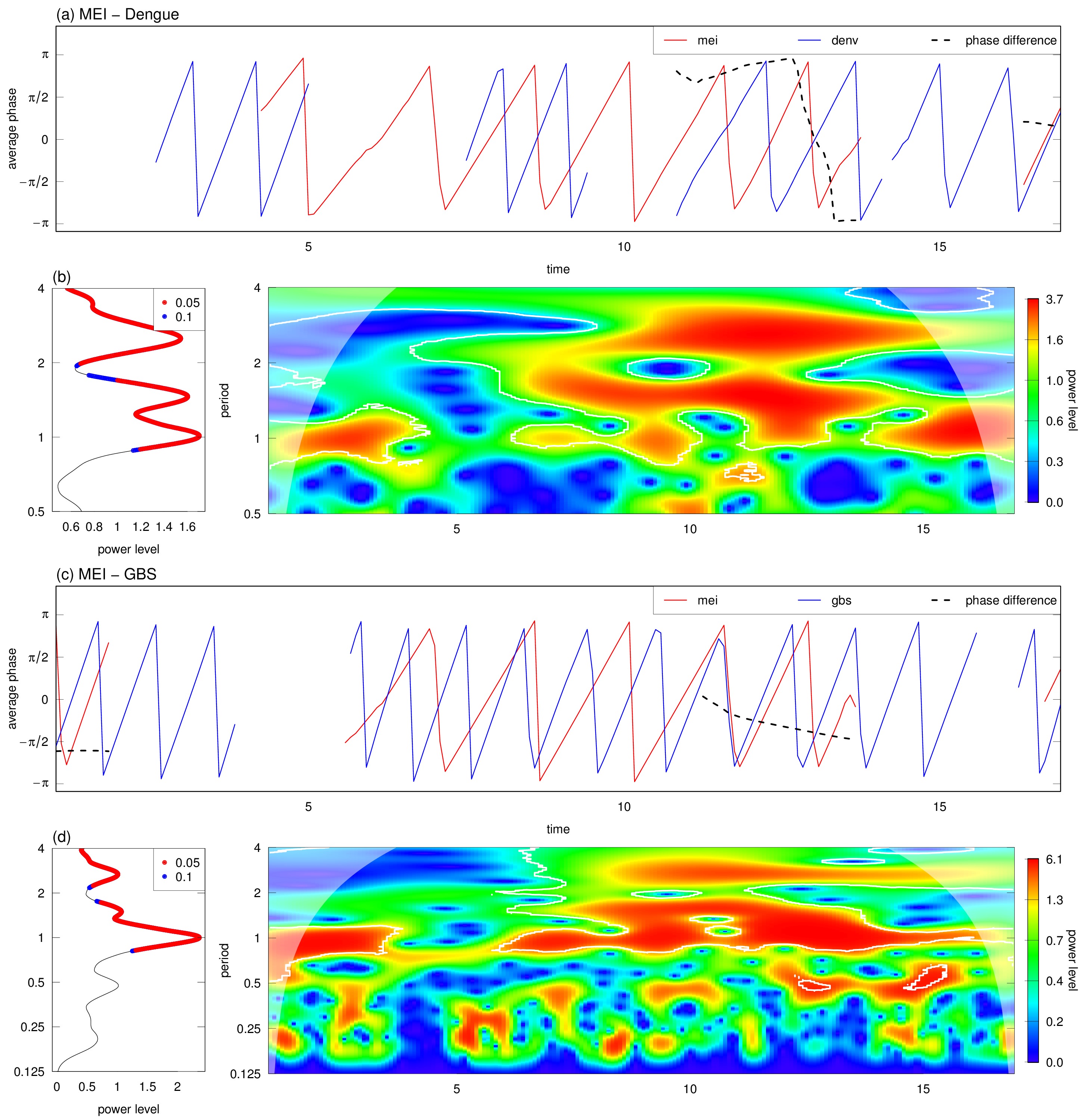}} 
\caption{Wavelet coherence and phase plots among MEI, dengue, and GBS from 2000 to 2016. (a) and (c) show phase analyses. Data are shown in red and blue, and black dashed line shows phase difference. (b) and (d) show the cross-wavelet average power level and wavelet coherence plots, which share the same colour schemes as in Fig. 4. All data are transformed by taking square roots.}
\label{fg:cross_wavelet_mei_dengue_and_gbs}
\end{figure}

\begin{figure}[!ht]
\centerline{\includegraphics[width=16.5cm,height = 13cm]{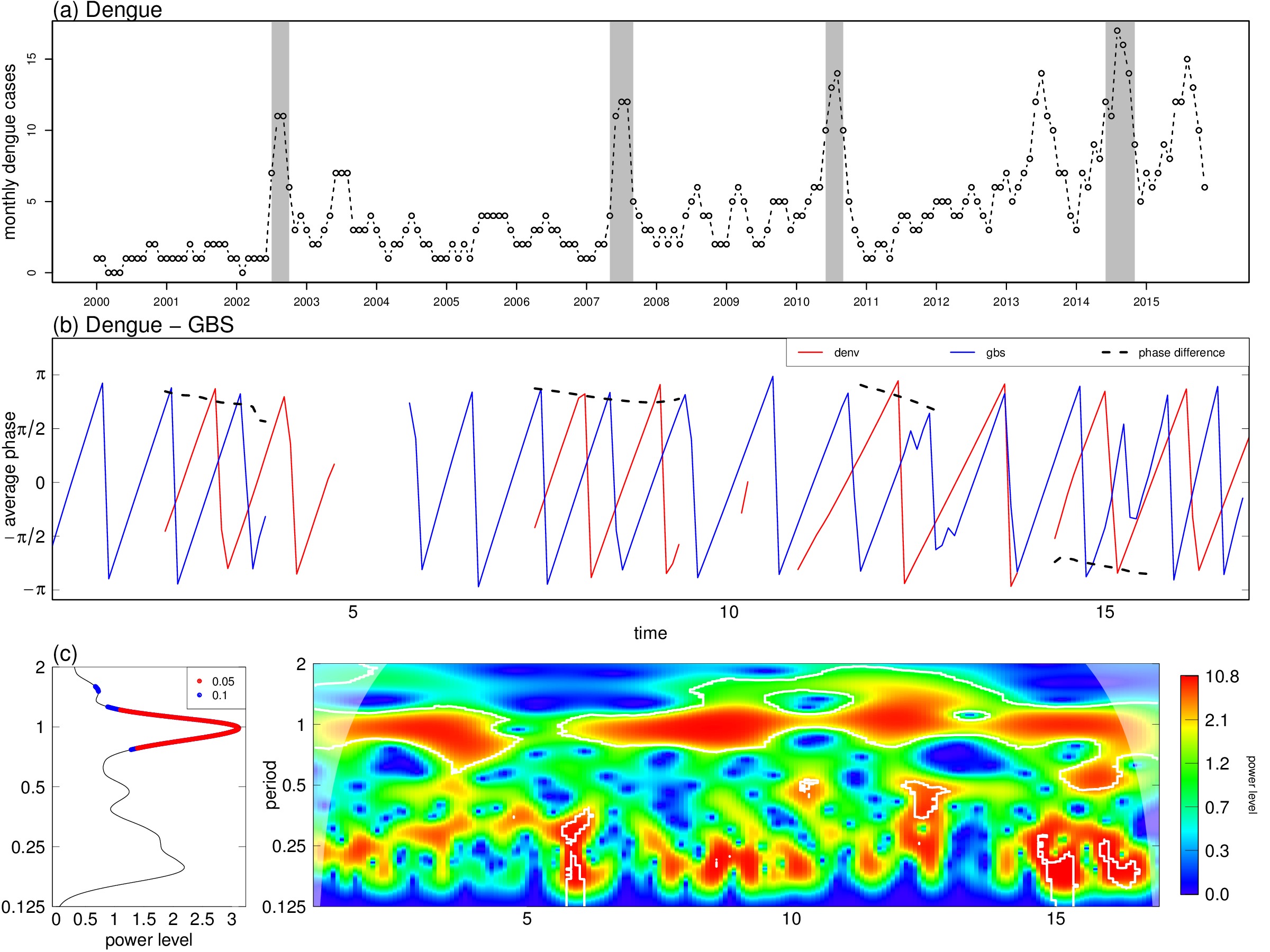}} 
\caption{Wavelet coherence and phase plots of dengue vs. GBS data from 2000-15 in Hong Kong. Panel (a) is dengue time series with peaks shaded in grey.
Panel (b) is phase plots of dengue vs. GBS, data are shown in red and blue, and black dashed line shows phase difference. Panel (c) show cross wavelet average power level and wavelet coherence plots of dengue vs. GBS, which share the same plot code as Fig.~\ref{fg:wavelet_mei_dengue_gbs}. The horizontal axis of 5, 10 and 15 represent year 2005, 2010 and 2015. All data are transformed by taking square roots.}
\label{fg:cross_wavelet_dengue_gbs}
\end{figure}

\section*{Discussion}
In this work, we reported increasing patterns of both GBS cases and imported dengue cases in Hong Kong, and investigated the possible mechanism behind said patterns. We found weak but statistically significant negative correlation between GBS and meteorological factors. Number of GBS cases was negatively correlated with both temperature and evaporation. Our findings are consistent with Webb et al's meta analyses that GBS cases were higher in winter than in summer \cite{webb2015seasonal}. The peak of dengue from 2013 to 2015 is largely consistent with that of MEI for the same period. MEI explained over 12\% of dengue's variations from Poisson regression models. Our results are consistent with previous studies \cite{tipayamongkholgul2009effects, hurtado2007short}.
Earlier clinical case studies reported dengue preceding GBS \cite{Qureshi, Chen, Kumar, Sulekha, Goncalves, Simon, Santos}. Our findings further showed that there is a cross-correlation between GBS and dengue cases. The increased magnitude of dengue outbreaks in Southern China could have played a role in the recent increases of GBS cases in Hong Kong.
Our wavelet results showed that dengue and MEI oscillated in one to two-year periodic band. Our findings are consistent with earlier findings conducted in Vietnam, Thailand, and Southeast Asian countries in general \cite{vanPanhuis+15, Cazelles2005nonstationary, Cuong}. However, Johansson reported no association between dengue and multiyear climate variability on a study conducted in Mexico, Thailand and Puerto Rico \cite{Johansson2009multiyear}. As no ZIKV outbreaks had been reported in Hong Kong as of to date, local GBS cases are unlikely to be triggered by ZIKV. Thus, it is justifiable to use dengue and MEI data as an early warning for GBS surveillance.
To our best knowledge, our study is the first to report on a possible non-stationary oscillation association between dengue and GBS. dengue reported cases displayed peaks in 2002, 2007, 2010 and 2014 respectively in Hong Kong, and phase plots of dengue and GBS indicated stronger coherency around those years. There are two major strengths in this study. First, although there had been several clinical case reports of dengue preceding GBS, we are novel to report on their association at the population level. Second, our wavelet analyses of GBS, dengue and MEI are well-suited to demonstrate the non-stationary oscillating association among them.
This study is limited by several factors . First, GBS has both infectious and non-infectious triggers, and we do not have information about the antecedent events of reported GBS cases. Second, most of the dengue cases are imported cases, but we did not consider the population's travel patterns and the source country of infected cases. Third, dengue reported cases could be an underestimate of the true number of dengue infections in Hong Kong, since dengue could be a mild non-specific febrile illness that is difficult to distinguish from other illnesses.
Our study has led to an improved understanding about the timing and relationship between MEI, GBS and dengue. Future studies should explore these disease patterns across a larger region scale to infer the mechanisms behind them. It would help to inform policymakers in designing appropriate prevention and control measures to combat these growing public health challenges.

\section*{Methods}
\subsection*{Data and Methods}
\subsubsection*{Epidemiological Data}
Monthly GBS cases from January 2000 to June 2016 and dengue cases from January 1999 to June 2016 are downloaded from the website of Centre for Health Protection in Hong Kong (http://www.chp.gov.hk). An infected patient who recently travelled to a dengue endemic country is considered an imported dengue case, otherwise it is considered locally-acquired.

\subsubsection*{Meteorological Data}
Meteorological data from January 1999 to June 2016 are downloaded from the website of Hong Kong Observatory (http://www.hko.gov.hk). After excluding missing data, we computed the median values of daily data in each month for further analyses. MEI data from January 1999 to June 2016 were downloaded from National Oceanic Atmospheric Administration's Earth System Research Laboratory (http://www.esrl.noaa.gov/psd/enso/mei/).

\subsubsection*{Statistical Analyses}
We computed the Spearman's Rank Correlation between monthly GBS cases with the monthly median values of daily meteorological factors from January 2000 to June 2016 in Hong Kong. We then introduced time lags and computed the cross-correlation coefficient between monthly dengue cases and monthly GBS cases. We estimated the association between dengue and MEI by applying Poisson Regression model.

\begin{equation}
\text{E}[\text{dengue}_{t+\tau} | \text{MEI}_{t}]=\lambda_{\text{dengue}_{t+\tau}}=\exp({\alpha+\beta \text{MEI}_t + \epsilon_{t+\tau}})
\label{regression_eqn}
\end{equation}
where where $\tau$ is the time lag with $\tau \in \{0,1,\dots,11\}$, $\lambda$ is the Poisson parameter of dengue cases number at time $(t+\tau)$, $\alpha$ and $\beta$ are the regression coefficients estimated by Maximum Likelihood approach and $\epsilon_{t+\tau}$ is the error term. The absolute value of coefficient of MEI, $\abs{\beta}$, can be interpreted as the non-linear association between MEI and dengue. The cross-correlation coefficients (ccf) are also used with different time lags.

Following previous works \cite{Cazelles2005nonstationary,Johansson2009multiyear,thai2010dengue}, we first transformed MEI, dengue and GBS data by taking square roots and then applied wavelet analyses to each set of these time series. Since MEI and the two diseases can be considered as ``natural signal" such that the Morlet wavelet, $\psi(\cdot)$, could be applied as the ``mother wavelet":

\begin{equation}
\psi(t) = \pi^{-\frac{1}{4}} ~ \exp[-i (2\pi f_0) t] ~ \exp(-\frac{1}{2} t^2)
\label{eq:morlet_wavelet}
\end{equation}
where $(2\pi f_0)$ is the relative frequency of the sine function. The wavelet transformation of our data is:

\begin{equation}
W_{\psi,x}(a,\kappa) = \int_{-\infty}^{\infty} x(t) \psi^*_{a,\kappa}(t) dt
\label{eq:wavelet_transform}
\end{equation}
where $W(\cdot)$ is the wavelet coefficient and it represents the contribution in transformation with $(a,\kappa)$ given, $a$ is the wavelet scale, and $\kappa$ represent different time positions and $x(t)$ denotes the time series (i.e. MEI, dengue and GBS).  $\psi^*_{a,\kappa}(\cdot)$ is the complex conjugation of the reformed ``mother wavelet", ie, Morlet wavelet. We then applied the cross wavelet analysis to quantify the association among each dataset.

Statistical software R is used for all statistical analyses (version Ri386 3.3.1).

\section*{Acknowledgements}
This study was supported by ShenZhen Science and Technology Innovation Project Grant(JCYJ20150402102135501) and Start-up Fund for New Recruits from Hong Kong Polytechnic University.

\section*{Author contributions statement}

TX, WX, ZS, CA and HD conceived the project. TX, ZS and HD analysed the results and wrote the draft. CA, YL and WX revised the article.
All authors reviewed the manuscript.

\section*{Competing interests}

The authors declare no competing financial interests.

\newpage
\section*{Appendix}
\subsection*{MEI Coefficient}
MEI Coefficients, $\beta$, of MA(3) and MA(5) in Poisson Regression model (eqn. \ref{regression_eqn})

\begin{table}[ht!]
\begin{center}
\caption{MEI coefficient, $\beta$, list of Poisson Regression model with MEI taken 3 Moving Average (MA).
Here, $***$ represents $p$-value under 0.0001, which is significant.}
\label{T:ma3_regression}
\begin{tabular}{r|rrcl}

\hline

Lag $\tau$& $\beta$ of MEI& AIC& $R^2$& signif\\

\hline

0&  0.1700& 1058.90& 0.0345& ***\\

1&  0.2095& 1045.20& 0.0507& ***\\

2&  0.2663& 1025.60& 0.0749& ***\\

3&  0.3172& 1004.20& 0.1073& ***\\

4&  0.3466& 989.08& 0.1224& ***\\

5&  0.3358& 985.40& 0.1131&  ***\\

6&  0.3130& 987.93& 0.0951&  ***\\

7&  0.2793& 991.02& 0.0754& ***\\

8&  0.2651& 988.23& 0.0868& ***\\

9&  0.2499& 985.30& 0.0619& ***\\

10&  0.2329& 984.21& 0.0544& ***\\

11&  0.2029& 985.24& 0.0419& ***\\

\hline

\end{tabular}
\end{center}
\end{table}

\begin{table}[ht!]
\begin{center}
\caption{MEI coefficient, $\beta$, list of Poisson Regression model with MEI taken 5 Moving Average (MA).
Here, $***$ represents $p$-value under 0.0001, which is significant.}
\label{T:ma5_regression}
\begin{tabular}{r|rrcl}

\hline

Lag $\tau$& $\beta$ of MEI& AIC& $R^2$& signif\\

\hline

0&  0.1677& 970.35& 0.0391& ***\\

1&  0.2019& 955.28& 0.0558& ***\\

2&  0.2400& 939.18& 0.0759& ***\\

3&  0.2758& 923.19& 0.0966& ***\\

4&  0.2963& 910.24& 0.1098& ***\\

5&  0.3028& 904.36& 0.1106&  ***\\

6&  0.2846& 903.50& 0.0970&  ***\\

7&  0.2711& 900.54& 0.0892& ***\\

8&  0.2518& 900.22& 0.0777& ***\\

9&  0.2322& 897.89& 0.0675& ***\\

10&  0.2128& 895.24& 0.0574& ***\\

11&  0.1963& 891.83& 0.0502& ***\\

\hline

\end{tabular}
\end{center}
\end{table}

\end{document}